\documentclass[12pt]{article}

\newcommand{\be}{\begin{equation}}
\newcommand{\ee}{\end{equation}}
\newcommand{\bi}[1]{\vspace{-3mm} \bibitem{#1}}

\usepackage{epsfig,amsmath,amssymb,graphics,graphicx} 

\begin{document}
\begin{center}

{\it Communications in Nonlinear Science and Numerical Simulation. Vol.20. (2015) 360-374.}


\vskip 3mm

{\bf \large Vector Calculus in
Non-Integer Dimensional Space and \\
its Applications to Fractal Media} \\

\vskip 7mm
{\bf \large Vasily E. Tarasov} \\
\vskip 3mm

{\it Skobeltsyn Institute of Nuclear Physics,\\ 
Lomonosov Moscow State University, Moscow 119991, Russia} \\
{E-mail: tarasov@theory.sinp.msu.ru} \\

\begin{abstract}
We suggest a generalization of vector calculus
for the case of non-integer dimensional space.
The first and second orders operations 
such as gradient, divergence, 
the scalar and vector Laplace operators 
for non-integer dimensional space are defined.
For simplification we consider scalar and vector fields
that are independent of angles.
We formulate a generalization of vector calculus
for rotationally covariant scalar and vector functions.
This generalization allows us to describe fractal media
and materials in the framework of continuum models
with non-integer dimensional space.
As examples of application of the suggested calculus,
we consider elasticity of fractal materials
(fractal hollow ball and fractal cylindrical pipe 
with pressure inside and outside),
steady distribution of heat in fractal media,
electric field of fractal charged cylinder.
We solve the correspondent equations 
for non-integer dimensional space models.
\end{abstract}

\end{center}


\section{Introduction}

In general we can assume that space and space-time dimensions 
are $D$, which need not be an integer. 
Non-integer dimensional spaces and method of 
dimensional continuation are initially emerged 
in statistical mechanics and quantum field theory. 
Non-integer dimension $D=4 -\varepsilon$ of space-time and $\varepsilon$-expansion are actively used in the theory of critical phenomena and phase transitions in statistical physics 
(for example, see \cite{WilsonFisher,WK1974}). 
Integration over non-integer dimensional spaces is used in
the dimensional regularization method as a powerful tool 
to obtain exact results without ultraviolet divergences 
in quantum field theory \cite{HV1972,Leibbrandt,Collins}. 
In quantum theory, the divergences are parameterized 
as quantities with coefficients $\varepsilon^{-1}=(4-D)^{-1}$, 
and then these divergences can be removed
by renormalization to obtain finite physical values.

The axioms for integrals in non-integer dimensional space are suggested by Wilson in \cite{Wilson}. 
These properties are natural and necessary in applications in different areas \cite{Collins}. 
Theory of integration in non-integer dimensional spaces has been suggested in \cite{Stillinger,Collins,PS2004}.
Stillinger introduces \cite{Stillinger} a mathematical basis of integration on spaces with non-integer dimensions. 
In \cite{Stillinger} has been suggested a generalization of 
the Laplace operator for non-integer dimensional spaces also. 
In the book by Collins \cite{Collins} the integration 
in non-integer dimensional spaces is formulated 
for  rotationally covariant functions. 
The product measure method, which is suggested in \cite{Svozil}, 
and the Stillinger's approach \cite{Stillinger}
are extended by Palmer and Stavrinou \cite{PS2004} to multiple variables and different degrees of confinement in orthogonal directions. In the paper \cite{PS2004} 
extensions of integration and scalar Laplace operator 
for non-integer dimensional spaces are suggested.

The scalar Laplace operators, which are suggested 
in \cite{Stillinger,PS2004} for non-integer dimensional spaces, have a wide application in physics and mechanics.
Non-integer dimensional space has successfully been used as an effective physical description. 
The Stillinger's form of Laplacian 
first applied by He \cite{XFHe1,XFHe2,XFHe3,XFHe4}, 
where the Schr\"odinger equation 
in non-integer dimensional space is used and
the real confining structure is replaced by 
an effective space, such that the measure of the anisotropy or confinement is given by the non-integer dimension. 
Non-integer dimensions is used by Thilagam
to describe stark shifts of excitonic complexes 
in quantum wells \cite{Thilagam1997a},
exciton-phonon interaction in fractional dimensional space
\cite{Thilagam1997b}, and blocking effects in quantum wells
\cite{Thilagam1999}.
The non-integer dimensional space approach is used
by Matos-Abiague 
\cite{MA2000a,MA2000b,MA2001a,MA2001b,MA2002a,MA2002b,MA2002c}.
to describe momentum operators for quantum systems and 
Bose-like oscillator in non-integer dimensional space,
the polaron effect in quantum wells. 
Quantum mechanical models with non-integer (fractional) 
dimensional space has been suggested
by Palmer and Stavrinou \cite{PS2004},
Lohe and Thilagam \cite{QM1}.
The non-integer dimensional space approach is used to describe
algebraic properties of Weyl-ordered polynomials 
for the momentum and position operators \cite{QM2,QM3} and
the correspondent coherent states \cite{QM4}.
The Stillinger's form of Laplacian has been applied to  
the Schr\"odinger equation in non-integer dimensional space 
by Eid, Muslih, Baleanu, Rabei in \cite{QM5,QM6},
Muslih and Agrawal \cite{Muslih2010,MA2012},
by Calcagni, Nardelli, Scalisi in \cite{QM8}.
The fractional Schr\"odinger equation with 
non-integer dimensions is considered
by Martins, Ribeiro, Evangelista, Silva, Lenzi in \cite{QM9}
and by Sandev, Petreska, Lenzi \cite{SPL2013}.
Recent progress in non-integer dimensional space approach
includes the description of
the scalar field on non-integer dimensional spaces
by Trinchero \cite{Tri2012},
the fractional diffusion equation in
non-integer dimensional space and its solutions
are suggested in \cite{LSTLRL}.
The gravity in fractional dimensional space is described by
Sadallah, Muslih, Baleanu in \cite{SMB,SM}, and
by Calcagni in \cite{Calcagni-1,Calcagni-2,Calcagni-3}.
The electromagnetic fields in non-integer dimensional space
are considered in \cite{MB2007,BGG2010,MSBR2010}
and \cite{ZMN2010,ZMN2011a,ZMN2011b,ZMN2011c,ZMN}. 


Unfortunately, the basic articles \cite{Stillinger,PS2004} 
proposed only the second order differential operators 
for scalar fields 
in the form of the scalar Laplacian in the non-integer
dimensional space. The first order operators such as  gradient, divergence, curl operators, and the vector Laplacian are not considered in \cite{Stillinger,PS2004}.
In the book \cite{ZMN} 
(see also \cite{ZMN2010,ZMN2011a,ZMN2011b,ZMN2011c}), 
the gradient, divergence, and curl operators
are suggested only as approximations of the square of 
the Palmer-Stavrinou form of Laplace operator.
Consideration only
the scalar Laplacian in non-integer dimensional space approach
greatly restricts us in application of continuum models
with non-integer dimensional space for fractal media and material.
For example, we cannot use the Stillinger's form of Laplacian
for displacement vector field ${\bf u}({\bf r},t)$ 
in elasticity and thermoelasticity theories. 
We cannot consider equations for 
the electric field ${\bf E}({\bf r},t)$ and 
the magnetic fields ${\bf B}({\bf r},t)$ 
for electromagnetic theory of fractal media
by using continuum models with 
non-integer dimensional space.

In this paper, we propose a vector calculus
for non-integer dimensional space and we define
the first and second orders differential vector
operations such as gradient, divergence, 
the scalar and vector Laplace operators 
for non-integer dimensional space.
For simplification we consider rotationally covariant 
scalar and vector functions that are independent of angles.
In order to derive the vector differential operators 
in non-integer dimensional space
we use the method of analytic continuation in dimension.
For this aim we get equations for 
these differential operators for
rotationally covariant functions in $\mathbb{R}^n$ 
for arbitrary integer $n$ to highlight the explicit 
relations with dimension $n$.
Then the vector differential operators  
for non-integer dimension $D$ 
are defined by analytic continuation 
from integer dimensions $n$ to non-integer $D$.
These proposed equations allows us 
to reduce $D$-dimensional vector 
differentiations to usual derivatives 
with respect to one variable $r=|{\bf r}|$. 
It allows us to reduce differential equations 
in non-integer dimensional space to
ordinary differential equations with respect to $r$. 
The proposed operators  
allows us to describe fractal materials and media
in the framework of continuum models
with non-integer dimensional spaces.
In order to give examples of the possible applications,
we consider continuum models of fractal media and materials 
in the elasticity theory 
in the heat theory, and in the theory of
electric fields.
The correspondent equations 
for non-integer dimensional space are solved.

\section{Fractal media}

The cornerstone of fractal media 
is the non-integer dimension \cite{Fractal2}
such as mass or charge dimensions \cite{CM,TarasovSpringer}. 
In general, fractal media and materials can be treated 
with three different approaches:
(1) Using the methods of "Analysis on fractals" \cite{Kugami,Strichartz-1,Strichartz-2,Harrison,Kumagai,DGV} it is possible to describe fractal materials;
(2) To describe fractal media we can apply 
fractional-integral continuous models suggested in 
\cite{PLA2005-1,AP2005-2,IJMPB2005-2,MPLB2005-1,TarasovSpringer} 
(see also \cite{FM-1,FM-2,FM-3,FM-4,FM-5,FM-6,FM-7}
and \cite{MOS-1,MOS-2,MOS-3,MOS-4,MOS-5,MOS-6,MOS-7}).
In this case we use integrations of non-integer orders and 
two different notions such as density of states and 
distribution function \cite{TarasovSpringer};  
(3) Fractal materials can be described by using
the theory of integration and differentiation for
a non-integer dimensional space \cite{Collins,Stillinger,PS2004}.

The first approach, which is based on the use of analysis 
on fractal sets, is the most stringent possible method 
to describe idealized fractal media.
Unfortunately, it has two lacks.
Firstly, a possibility of application of 
the analysis on fractals to solve differential equations 
for real problems of fractal material is very limited 
due to weak development of this area of mathematics 
at this moment. 
Secondly, fractal materials and media cannot 
be described as fractals.
The main property of the fractal is non-integer 
Hausdorff dimension \cite{Fractal1} 
that should be observed on all scales. 
The fractal structure of real media 
cannot be observed on all scales
from the infinitely small to the infinitely large sizes.
Materials may have a fractal structure only
for scales from the characteristic size 
of atoms or molecules of fractal media
up to size of investigated sample of material.

The second approach, which is based on 
the use of fractional integration
in integer dimensional spaces, 
can give adequate models to describe fractal media.
The main disadvantage of the fractional-integral continuum models
is the existence of various types of fractional integrals, 
which led to the arbitrariness in the choice of 
the correspondent densities of states.


In this paper, we consider the third approach used 
the non-integer dimensional spaces.
One of the advantages of this approach is a possibility 
to avoid the arbitrary choice of densities of states.
In addition, we also suggest a generalization of the vector 
calculus to the case of non-integer dimensional space.
It allows us to use different continuum models
of fractal media and materials
in the framework of non-integer dimensional space approach.

Real fractal materials can be characterized by 
the relation between the mass $M_D(W)$ of a ball region $W$ 
of fractal medium, and the radius $R$ of this ball in the form
$M_D(W) = M_0 (R/ R_0)^D$, $R/R_0 \gg 1$,
where $R_0$ is the characteristic size of fractal medium
such as a minimal scale of self-similarity 
of a considered fractal medium. 
The number $D$ is called the mass dimension. 
The parameter $D$, does not depend on the shape 
of the region $W$, or on whether the packing of 
sphere of radius $R_0$ is close packing,
a random packing or a porous packing 
with a uniform distribution of holes.
As a result, fractal materials can be considered 
as media with non-integer mass dimensions. 
Although, the non-integer dimension does not reflect 
completely the geometric and dynamic properties 
of the fractal media, 
it nevertheless allows us to get 
important conclusions about the behavior of these media.  
As it will be shown in the next section, 
the power law $M_D(W) \sim R^{D}$ can be naturally 
derived by using the integration over 
non-integer dimensional space, where
the space dimension is equal to
the mass dimension of fractal media.


In order to describe fractal media by 
continuum models with non-integer dimensional space,
we should use the concepts of 
density of states $c_n(D,{\bf r})$
and distribution function $\rho({\bf r})$.
The density of states describes 
how closely packed permitted places (states) 
in the space $\mathbb{R}^n$, 
where the fractal medium is distributed.
The expression $c_n(D,{\bf r})dV_n$ is equal to
the number of permitted places (states) 
between $V_n$ and $V_n +dV_n$ in $\mathbb{R}^n$. 
The distribution function describes 
a distribution of physical values 
such as mass, electric charge, number of particles 
on a set of permitted places (possible states).
In general, the concepts of density of states and 
distribution function are different.
We cannot reduce all properties of the fractal media  
to the distribution function only, and
we should use the concepts of density of states 
to characterize how closely packet permitted states
places in the media.


The most important property of fractal medium
is the fractality, which means that 
the mass $M_D(W)$ of this medium 
in any region $W \subset \mathbb{R}^n$ increases more slowly 
than the $n$-dimensional volume $V_n(W)$ of this region.
For the ball region of the fractal medium, 
this property can be described by 
the power law $M_D(W) \sim R^{D}$, where $R$ is the radius 
of the ball, and $D$ is the mass dimension. 

Another important property of some fractal media is homogeneity.
Fractal medium is called homogeneous 
if the power law $M_D(W) \sim R^{D}$ does not depend on 
the translation of the region. 
The homogeneity property of the fractal medium means 
that two regions $W_1$ and $W_2$ 
with the equal volumes $V_n(W_1)=V_n(W_2)$ 
have equal masses $M_D(W_1)=M_D(W_2)$. 

To adequately describe the fractal media by 
continuum models with non-integer dimensional spaces, 
the following two requirements must be satisfied.

\begin{itemize} 
\item 
In the continuum models 
the mass density of homogeneous fractal medium should be described by the constant 
distribution function $\rho({\bf r})=\rho_0=const$. 
Then equations with constant density should
describe the homogeneous media, i.e., 
the conditions $\rho({\bf r})=const$ and $V_n(W_1)=V_n(W_2)$ 
should lead to the relation $M_D(W_1)=M_D(W_2)$. 

\item 
In the continuum models
the mass of the ball region $W$ of fractal homogeneous medium 
should be described by a power law relation $M \sim R^{D}$, 
where $0<D<3$, and $R$ is the radius of the ball. 
Then the conditions 
$V_n(W_1)= \lambda^n V_n(W_2)$ and $\rho({\bf r})=const$,
should lead to the relation $M_D(W_1)=\lambda^D M_D(W_2)$. 
\end{itemize}

These requirements cannot be realized
if the mass of fractal medium is described by integration
of integer order over the integer dimensional space
without using the concept of density of states $c_n(D,{\bf r})$. 
In order to realize these requirements 
we propose to use the integration and differentiation
in non-integer dimensional spaces.
In this case we can use the equation
\be \label{1-MW3} 
dM_D(W)= \rho({\bf r}) \, d V_D ({\bf r},n) , 
\ee
where 
$\rho({\bf r})$ is a distribution function, and
the density of states 
$c_n(D,{\bf r})$ in $\mathbb{R}^n$ is chosen such that
\[ d V_D ({\bf r},n) = c_n(D,{\bf r}) \, dV_n  \]
describes the number of permitted states in $dV_n$.
For different values of $n \in \{1,2,3\}$ 
we can use the notations 
\be
dV_D=c_3(D,{\bf r}) \, dV_3 ,  \quad
dS_d=c_2(d,{\bf r}) \, dS_2, \quad 
dl_{\beta}=c_1(\beta,{\bf r})\, dl_1
\ee
to describe fractal media in the spaces $\mathbb{R}^n$, 
where these media are distributed.
For simplification, we also use the notation $d^D {\bf r}$ 
instead of $ d V_D ({\bf r},n)$.
The form of function $c_n(D,{\bf r})$ is defined by the 
properties of considered fractal medium. 
The symmetry of the density of states $c_n(D,{\bf r})$  
is dictated by the symmetry properties of 
the described fractal medium, but in any cases
it should be a function of power-law type 
to adequately reflect a scaling property (fractality) 
of described fractal medium. 
To simplify the analysis in this article we will consider only isotropic fractal media with densities of states 
that are independent of angles.



In the continuum models of fractal media,
it is convenient to work in the dimensionless space variables
$x/R_0 \to x$, $y/R_0 \to x$, $z/R_0 \to x$, 
${\bf r}/R_0 \to {\bf r}$,
that yields dimensionless integration and dimensionless differentiation in non-integer dimensional space.
In this case the physical and mechanical quantities of 
fractal media have correct physical dimensions.


\section{Integration over non-integer dimensional space}

The integral for all non-integer values of $D$ 
is defined by continuation in $D$ \cite{Leibbrandt,Collins}.   
Let us give properties must we impose 
on a functional of $f({\bf r})$ 
in order to regard it as $D$-dimensional integration. 
The following properties or axioms \cite{Wilson}
for integrals in $D$-dimensional space   
are natural and necessary in applications \cite{Collins}: 

\begin{enumerate} 

\item 
Linearity: 
\be \label{Ax-1}
\int \Bigl( af_1({\bf r})+bf_2({\bf r}) \Bigr) \, d^D {\bf r} =
a \int f_1({\bf r}) \, d^D r + b \int f_2({\bf r}) \, d^D {\bf r} ,
\ee
where $a$ and $b$ are arbitrary real numbers, and
$d^D {\bf r}= d V_D ({\bf r},n)$ represents 
the volume element in the non-integer dimensional space..

\item 
Translational invariance:
\be \label{Ax-2}
\int f({\bf r}+{\bf r}_0) \, d^D {\bf r} = 
\int f({\bf r}) \, d^D {\bf r} 
\ee
for any vector ${\bf r}_0$.

\item 
Scaling property: 
\be \label{Ax-3}
\int f(\lambda {\bf r} ) \, d^D {\bf r} = 
\lambda^{-D} \int f({\bf r}) \, d^D {\bf r}
\ee
for any positive $\lambda$.

\end{enumerate}

Linearity is true of any integration, 
while translation and rotation invariance 
are basic properties of a Euclidean space.
The scaling property embodies the $D$-dimensionality. 
Not only are the above three axioms necessary, 
but they also ensure that integration is unique, 
aside from an overall normalization \cite{Wilson}. 

These properties must be imposed on a functional of $f({\bf r})$
in order to regard it as $D$-dimensional integrations \cite{Collins}.
These properties are natural and are necessary 
in application of dimensional regularization
to quantum field theory (see section 4 in \cite{Collins}).


A function $f({\bf r})$ that we integrate could 
in principle be any function of the components 
of its vector argument ${\bf r}$. 
However, we do not, a priori, know the meaning 
of the components of, say, a vector in non-integer dimensions. 
In this paper, we will work 
with rotationally covariant functions. 
So we will assume that $f$ is a scalar or vector function 
only of scalar products of vectors or of length of vectors. 
For example, in the elasticity theory, we consider the case,
where the displacement vector ${\bf u} ({\bf r})$, 
is independent of the angles
${\bf u} ({\bf r}) = {\bf u}(r)$, where $r=|{\bf r}|$.

The integral defined in equation (\ref{1-dimregint}) 
satisfies the properties (\ref{Ax-1}) - (\ref{Ax-3}). 


The $D$-dimensional integration for
scalar functions $f({\bf r})=f(|{\bf r}|)$ can be defined 
in terms of ordinary integration by the expression
\be \label{1-dimregint}
\int d^D {\bf r} \ f({\bf r}) 
= \int_{\Omega_{D-1}} d\Omega_{D-1} 
\int^{\infty}_0 dr \; r^{D-1} \; f(r) ,
\ee
where we can use
\be \label{SD-1}
\int_{\Omega_{D-1}} d\Omega_{D-1} = 
\frac{2 \pi^{D/2}}{\Gamma(D/2)} =S_{D-1}. \ee
For integer $D=n$, equation (\ref{SD-1}) gives
the well-known area $S_{n-1}$ 
of $(n-1)$-sphere with unit radius.

As a result, we have \cite{Collins} 
the explicit definition of the continuation of integration
from integer $n$ to arbitrary fractional $D$ in the form
\be \label{1-dim-reg} 
\int d^D {\bf r} \ f(|{\bf r}|) = \frac{2 \pi^{D/2}}{\Gamma(D/2)} 
\int^{\infty}_0 dr \; r^{D-1} \; f(r) . \ee
This equation reduced $D$-dimensional integration 
to ordinary integration.
Therefore the linearity and translation invariance 
follow from linearity and translation invariance 
of ordinary integration. 
The scaling and rotation covariance are explicit properties of the definition.


Let us give some examples of an application 
of equation (\ref{1-dim-reg}). 
For the function
\be \label{f-example}
f({\bf r}^2)= \frac{{\bf r}^2+a}{{\bf r}^2+b} , \ee
where $a$ and $b$ are real numbers. 
The integral for (\ref{f-example}) can be explicitly computed
\be
\int d^D {\bf r} \, \frac{r^2+a}{r^2+b} =
(\pi b)^{D/2} \, (a/b-1) \, \Gamma(1-D/2) .
\ee
The other example is the integral
\be
\int d^D {\bf r} \, \frac{r^{2 \alpha}}{(r^2+a^2)^{\beta}}= 
\frac{\Gamma(\alpha+D/2) \, 
\Gamma(\beta-\alpha-D/2)}{\Gamma(D/2) \Gamma(\beta)} \,
\pi^{D/2} \, a^{D+2\alpha -2\beta} ,
\ee
where $r=|{\bf r}|$.

The mass of material in $W$ is described by the integral 
\be \label{1-MW3} 
M_D(W) = \int_{W} \rho({\bf r}) \, d^D {\bf r} ,
\ee
where ${\bf r}$ is dimensionless vector variable. 
For a ball with radius $R$ and the density 
$\rho({\bf r})=\rho_0 =\operatorname{const}$, we get 
the mass is defined by
\be \label{M-D}
M_D(W) = \rho_0 \, V_D =
\frac{\pi^{D/2} \, \rho_0}{\Gamma(D/2+1)} \, R^{D} .
\ee
This equation defines the mass of the 
fractal homogeneous ball with volume $V_D$. 
For $D=3$, equation (\ref{M-D}) gives the well-known 
equation for mass of non-fractal ball
$M_3 =(4 \rho_0 \pi/3) R^3$ because 
$\Gamma(3/2) =\sqrt{\pi}/2$ and $\Gamma(z+1)=z \, \Gamma(z)$. \\

\section{Divergence for non-integer dimensional space}

Let us consider hollow ball $B_D(R_1;R_2)$  
with internal radius $R_1$ and external radius $R_2$
in non-integer dimensional space.
The boundary $\partial B_D(R_1;R_2)$ of this ball 
consists of two $(D-1)$-dimensional spheres 
$S_{D-1}(R_1)$ and $S_{D-1}(R_2)$.

We assume that the vector field ${\bf u}({\bf r})$ 
is radially directed and ${\bf u}({\bf r})$ 
is not dependent on the angles, i.e., 
\be \label{ure-1}
{\bf u} ({\bf r}) =u_r(r) \, {\bf e}_r ,
\ee
where ${\bf e}_r = {\bf r} / |{\bf r}|$ and $r=|{\bf r}|$.
We can defined a flux of the vector field ${\bf u}(r)$ 
across a surface $\partial B_D(R_1;R_2)$ 
by using the integration 
in non-integer dimensional space suggested in \cite{Collins}.
A flux of the vector field ${\bf u}(r)$ across 
a $(D-1)$-dimensional surface $S_{D-1}$ is the integral
\be
\Phi_{\bf u} (S_{D-1})= 
\int_{S_{D-1}} ({\bf u}, d {\bf S}_{D-1}) ,
\ee
where $d {\bf S}_{D-1} ={\bf e}_r \, d S_{D-1}$.

The volume of $B_D(R_1;R_2)$ is equal to
\be
V(B_D(R_1;R_2)) = \frac{\pi^{D/2}}{\Gamma(D/2+1)} 
\, (R^D_2 -R^D_1) .
\ee  

An exact expression for dependence of 
the divergence operator on the non-integer dimension $D$ 
and the vector field ${\bf u} ({\bf r}) =u_r(r) \, {\bf e}_r$ 
can be derived by the equation
\be
\operatorname{Div}^{D}_{r} {\bf u} = \lim_{V(B_D(R_1;R_2)) \to 0}
\frac{ \Phi_{\bf u} (\partial B_D(R_1;R_2)) }{V(B_D(R_1;R_2))} .
\ee
Here the flux of the vector field can be represented by
\be
 \Phi_{\bf u} (\partial B_D(R_1;R_2))  =
\int_{S_{D-1}(R_2)} ({\bf u}, d {\bf S}_{D-1}) -
\int_{S_{D-1}(R_1)} ({\bf u}, d {\bf S}_{D-1}) .
\ee
Using (\ref{ure-1}), we get
\[
\int_{\partial B_D(R_1;R_2)} ({\bf u}, d {\bf S}_{D-1}) =
\int_{S_{D-1}(R_2)} ({\bf u}, d {\bf S}_{D-1}) -
\int_{S_{D-1}(R_1)} ({\bf u}, d {\bf S}_{D-1}) =
\]
\[
= u_r(R_2) \, \int_{S_{D-1}(R_2)} d S_{D-1} -
u_r(R_1) \, \int_{S_{D-1}(R_1)}  d S_{D-1} =
\]
\be
= \frac{2 \, \pi^{D/2}}{\Gamma(D/2)} 
\, \Bigl( u_r(R_2) \, R^{D-1}_2 - u_r(R_1) \, R^{D-1}_1 \Bigr) ,
\ee
where $S_{D-1} = ({\bf e}_r, {\bf S}_{D-1})$.

For $D=3$, we have 
$\Gamma (3/2)= (1/2) \, \Gamma (1/2) = \sqrt{\pi}/2$, and
\be
\int_{\partial B_D(R_1;R_2)} ({\bf u}, d {\bf S}_{D-1}) =
4 \pi \, \Bigl( u_r(R_2) \, R^2_2 - u_r(R_1) \, R^2_1 \Bigr) .
\ee

For an infinitely thin hollow ball (thin spherical shell) 
with $R_1=r$ and $R_2=r+ \Delta r$, we can use
\be
\operatorname{Div}^{D}_{r} {\bf u} =
\lim_{ \Delta r \to 0}
\frac{ \Phi_{\bf u} (\partial B_D(r;r+\Delta r)) }{ 
V(B_D(r;r+\Delta r))} 
\ee
to derive an expression for the divergence for our case.

The volume of the ball $B_D(r;r+\Delta r)$ is 
\[
V(B_D(r;r+\Delta r)) = \frac{\pi^{D/2}}{\Gamma(D/2+1)} 
\, \Bigl( (r+\Delta r)^D -r^D \Bigr) =
\]
\be \label{Vol-1}
= \frac{\pi^{D/2}}{\Gamma(D/2+1)} 
\, \Bigl( D \, r^{D-1} \, \Delta r +O((\Delta r)^2) \Bigr) = 
\frac{2\pi^{D/2}}{\Gamma(D/2)} 
\, \Bigl( r^{D-1} \, \Delta r +O((\Delta r)^2) \Bigr) .
\ee  

The flux of the vector field ${\bf u}=u_r(r) \, {\bf e}_r$ 
across a surface $\partial B_D(R_1;R_2)$ is given by
\[ 
\Phi_{\bf u} (\partial B_D(r;r+\Delta r)) =
\int_{S_{D-1}(r+\Delta r)} ({\bf u}, d {\bf S}_{D-1}) -
\int_{S_{D-1}(r)} ({\bf u}, d {\bf S}_{D-1}) = \]
\[
= \frac{2 \, \pi^{D/2}}{\Gamma(D/2)} 
\, \Bigl( u_r(r+\Delta r) \, (r+\Delta r)^{D-1}
- u_r(r) \, r^{D-1} \Bigr) =
\]
\[
= \frac{2 \, \pi^{D/2}}{\Gamma(D/2)} 
\, \Bigl( \Bigl[ u_r(r) +\frac{\partial u_r(r)}{\partial r} \Delta r
+O((\Delta r)^2) \Bigr] \, \Bigl[ r^{D-1} + 
(D-1) \, r^{D-2} \, \Delta r +O((\Delta r)^2) \Bigr] 
- u_r(r) \, r^{D-1} \Bigr) =
\]
\be \label{Flux-1}
= \frac{2 \, \pi^{D/2}}{\Gamma(D/2)} 
\, \Bigl( r^{D-1} \, \frac{\partial u_r(r)}{\partial r} +
(D-1) \, r^{D-2} \, u_r (r) +O(\Delta r) \Bigr) \Delta r .
\ee

Using (\ref{Vol-1}) and (\ref{Flux-1}), we get
\be
\frac{\Phi_{\bf u} (\partial B_D(r;r+\Delta r)) }{ 
V(B_D(r;r+\Delta r))}  =
\frac{\partial u_r(r)}{\partial r} + \frac{D-1}{r} \, u_r(r) 
+O(\Delta r)  .
\ee
As a result, the divergence 
for the vector field ${\bf u}=u_r(r)\, {\bf e}_r$ 
in non-integer dimensional space 
has the form
\be
\operatorname{Div}^{D}_{r} {\bf u} =
\lim_{ \Delta r \to 0} 
\frac{\Phi_{\bf u} (\partial B_D(r;r+\Delta r)) }{ 
V(B_D(r;r+\Delta r))}  =
\frac{\partial u_r(r)}{\partial r} + \frac{D-1}{r} \, u_r(r) .
\ee

The Gauss theorem in non-integer dimensional space 
can be written in the form
\be
\int_{B_D(R_1;R_2)} \operatorname{Div}^{D}_{r} {\bf u} \ 
d^D {\bf r} =
\int_{\partial B_D(R_1;R_2)} ({\bf u}, d {\bf S}_{D-1}) ,
\ee
where we assume that
the dimension $D$ of the region of fractal materials 
and the dimension $d$ of boundary of this region
are related by the equation $d = D-1$.

\section{Vector differential operators 
in non-integer dimensional space}

We would like to derive equations for 
vector differential operators 
in non-integer dimensional space.
For this aim we should have equations for 
these differential operators for
rotationally covariant functions
in the spherical coordinates in $\mathbb{R}^n$ 
for arbitrary $n$ to highlight the explicit 
relations with dimension $n$.
Then the vector differential operators  
for non-integer dimension $D$ 
can be defined by analytic continuation in 
dimension from integer $n$ to non-integer $D$. 

To simplify we will consider only scalar 
fields $\varphi$ and vector fields ${\bf u}$ 
that are independent of angles 
\[ \varphi({\bf r}) =\varphi(r) ,
\quad {\bf u}({\bf r})={\bf u} (r) = u_r \, {\bf e}_r , \]
where ${\bf e}_r={\bf r} /r$, $r=|{\bf r}|$ and $u_r=u_r(r)$
is the radial component of ${\bf u}$.
We will work with rotationally covariant functions only. 
This simplification is analogous to the simplification
for definition of integration over non-integer dimensional space
suggested in \cite{Collins}.


Let us give equations for differential operators 
for functions ${\bf u}=u_r(r)\, {\bf e}_r$
and $\varphi = \varphi(r)$
in the spherical coordinates in $\mathbb{R}^n$ 
for arbitrary $n$.

The divergence in integer dimensional space $\mathbb{R}^n$ 
for the vector field ${\bf u}=u(r) \, {\bf e}_r$ is
\be \label{Div-n}
\operatorname{div} {\bf u} = 
\frac{\partial u_r}{\partial r} + \frac{n-1}{r} \, u_r .
\ee

The gradient in integer dimensional space $\mathbb{R}^n$ 
for the scalar field $\varphi=\varphi (r)$ is
\be \label{Grad-n}
\operatorname{grad} \varphi = 
\frac{\partial \varphi}{\partial r} \, {\bf e}_r .
\ee

The scalar Laplacian in integer dimensional space $\mathbb{R}^n$ 
for the scalar field $\varphi=\varphi (r)$ is
\be \label{S-Delta-n}
\Delta \varphi (r)= 
\operatorname{div} \operatorname{grad} \varphi (r)= 
\frac{\partial^2 \varphi}{\partial r^2} + \frac{n-1}{r} \, 
\frac{\partial \varphi}{\partial r} .
\ee

The vector Laplacian \cite{VLap}
in integer dimensional space $\mathbb{R}^n$ 
for the vector field ${\bf u}=u(r) \, {\bf e}_r$ is
\be \label{V-Delta-n}
\Delta {\bf u}(r) = 
\operatorname{grad} \operatorname{div} {\bf u}(r) = 
\Bigl(
\frac{\partial^2 u_r}{\partial r^2} + \frac{n-1}{r} \, 
\frac{\partial u_r}{\partial r}  -  \frac{n-1}{r^2} \, u_r 
\Bigr) \, {\bf e}_r .
\ee


As a result, we have equations of  
differential operators in $\mathbb{R}^n$ for continuation
from integer $n$ to arbitrary non-integer $D$ 
in the following forms.

The divergence in non-integer dimensional space 
for the vector field ${\bf u}={\bf u}(r)$ is
\be \label{Div-D}
\operatorname{Div}^{D}_{r} {\bf u} = 
\frac{\partial u_r}{\partial r} + \frac{D-1}{r} \, u_r .
\ee

The gradient in non-integer dimensional space 
for the scalar field $\varphi=\varphi (r)$ is
\be \label{Grad-D}
\operatorname{Grad}^{D}_r \varphi = 
\frac{\partial \varphi}{\partial r} \, {\bf e}_r .
\ee

The curl operator for the vector field ${\bf u}={\bf u}(r)$
is equal to zero
\be \label{Curl-D}
\operatorname{Curl}^{D}_r {\bf u} = 0 .
\ee

The scalar Laplacian in non-integer dimensional space 
for the scalar field $\varphi=\varphi (r)$ is
\be \label{S-Delta-D}
^S\Delta^{D}_r \varphi= 
\operatorname{Div}^{D}_r \operatorname{Grad}^{D}_{r} \varphi =
\frac{\partial^2 \varphi}{\partial r^2} + \frac{D-1}{r} \, 
\frac{\partial \varphi}{\partial r} .
\ee

The vector Laplacian in non-integer dimensional space 
for the vector field ${\bf u}=u(r) \, {\bf e}_r$ is
\be \label{V-Delta-D}
^V\Delta^{D}_r {\bf u} = 
\operatorname{Grad}^{D}_r \operatorname{Div}^{D}_{r} {\bf u} =
\Bigl(
\frac{\partial^2 u_r}{\partial r^2} + \frac{D-1}{r} \, 
\frac{\partial u_r}{\partial r}  -  \frac{D-1}{r^2} \, u_r 
\Bigr) \, {\bf e}_r.
\ee


Let us consider a case of 
axial symmetry of the fractal material, where
the fields $\varphi(r)$ and ${\bf u}(r)=u_r(r) \, {\bf e}_r$
are also axially symmetric.
Let $Z$-axis be directed along the axis of symmetry. 
Therefore it is convenient to use 
a cylindrical coordinate system.
Equations for differential vector operations
for cylindrical symmetry case have the following forms.

The divergence in non-integer dimensional space 
for the vector field ${\bf u}={\bf u}(r)$ is
\be \label{Div-DC}
\operatorname{Div}^{D}_{r} {\bf u} = 
\frac{\partial u_r}{\partial r} + \frac{D-2}{r} \, u_r .
\ee

The gradient in non-integer dimensional space 
for the scalar field $\varphi=\varphi (r)$ is
\be \label{Grad-DC}
\operatorname{Grad}^{D}_r \varphi = \frac{\partial \varphi}{\partial r} \, {\bf e}_r .
\ee

The scalar Laplacian in non-integer dimensional space 
for the scalar field $\varphi=\varphi (r)$ is
\be \label{S-Delta-DC}
^S\Delta^{D}_r \varphi= 
\operatorname{Div}^{D}_r \operatorname{Grad}^{D}_{r} \varphi =
\frac{\partial^2 \varphi}{\partial r^2} + \frac{D-2}{r} \, 
\frac{\partial \varphi}{\partial r} .
\ee

The vector Laplacian in non-integer dimensional space 
for the vector field ${\bf u}=u(r) \, {\bf e}_r$ is
\be \label{V-Delta-DC}
^V\Delta^{D}_r {\bf u} = 
\operatorname{Grad}^{D}_r \operatorname{Div}^{D}_{r} {\bf u} =
\Bigl(
\frac{\partial^2 u_r}{\partial r^2} + \frac{D-2}{r} \, 
\frac{\partial u_r}{\partial r}  -  \frac{D-2}{r^2} \, u_r 
\Bigr) \, {\bf e}_r.
\ee


For $D=3$ equations (\ref{Div-D}) - (\ref{V-Delta-DC})
give the well-known expressions for the gradient,
divergence, scalar Laplacian and vector Laplacian
in $\mathbb{R}^3$ for fields
$\varphi=\varphi(r)$ and ${\bf u}(r)=u_r(r) \, {\bf e}_r$.

It is easy to generalize these equations for
the case $\varphi=\varphi(r,z)$ and 
${\bf u}(r,z)=u_r(r,z) \, {\bf e}_r+ u_r(r,z) \, {\bf e}_z$.
In this case the curl operator for ${\bf u}(r,z)$
is different from zero, and
\be \label{Curl-DC}
\operatorname{Curl}^{D}_{r,z} {\bf u} = 
\left( \frac{\partial u_r}{\partial z} -
\frac{\partial u_z}{\partial r} \right) \, {\bf e}_r .
\ee


The suggested operators allow us to reduce 
$D$-dimensional vector differentiations 
(\ref{Div-D}) - (\ref{V-Delta-D}) and 
(\ref{Div-DC}) - (\ref{V-Delta-DC})
to derivatives with respect to $r=|{\bf r}|$. 
It allows us to reduce partial differential equations for
fields in non-integer dimensional space to
ordinary differential equations with respect to $r$.

\section{Stillinger's Laplacian for non-integer dimensional space}


In the paper \cite{Stillinger}, 
the integration in a non-integer dimensional space 
is described by using the equation
\be \label{NI-2}
\int_{R^D} d^D {\bf r} \, \varphi ({\bf r}) =
\frac{2 \pi^{(D-1)/2}}{\Gamma((D-1)/2)}
\int^{\infty}_0 dr \, r^{D-1} \,
\int^{\pi}_0 d \theta \, \varphi (r,\theta) \, sin^{D-2}\theta \, ,
\ee
where $d^D {\bf r}= d V_D ({\bf r},n)$ 
represents the volume element
in the non-integer dimensional space.
Using (\ref{NI-2}) with $\varphi (r,\theta)=1$, and
\be \int^{\pi}_0 d \theta \, sin^{D-2}\theta = 
\frac{\pi^{1/2}\, \Gamma (D/2-1)}{\Gamma(D/2)} , \ee
we get
\be
V_D = \frac{\pi^{D/2}}{\Gamma(D/2+1)} \, R^{D} ,
\ee
which is the volume of a $D$-dimensional ball with radius $R$.

Using the analytic continuation of Gaussian integrals
the scalar Laplace operator for non-integer dimensional space 
has been suggested in \cite{Stillinger}.
For a function $ \varphi = \varphi (r,\theta)$ 
of radial distance $r$ and related angle $\theta$ 
measured relative to an axis passing through the origin,
the Laplacian in non-integer dimensional space
proposed by Stillinger \cite{Stillinger} is
\be \label{NI-1}
^{St}\Delta^{D} = \frac{1}{r^{D-1}} \frac{\partial}{\partial r} \left( r^{D-1} \,\frac{\partial}{\partial r} \right) +
\frac{1}{r^2 \, \sin^{D-2} \theta} 
\frac{\partial}{\partial \theta} 
\left(  \sin^{D-2} \theta 
\frac{\partial}{\partial \theta} \right) ,
\ee
where $D$ is the dimension of space ($0<D <3$),
and the variables $r \ge 0$, $0\le \theta \le \pi$.
Note that
$(\, ^{St}\Delta^{D} )^2 \ne \, ^{St}\Delta^{2D}$.
If the function depends on radial distance $r$ 
only ($\varphi =\varphi (r)$), then
\be \label{NI-R}
^{St}\Delta^{D} \varphi (r)= \frac{1}{r^{D-1}} \frac{\partial}{\partial r} \left( r^{D-1} \,\frac{\partial \varphi (r)}{\partial r} \right) =
\frac{\partial^2 \varphi (r)}{\partial r^2} +
\frac{D-1}{r} \, \frac{\partial \varphi (r)}{\partial r} .
\ee
It is easy to see that 
the Stillinger's form of Laplacian $\, ^{St}\Delta^{D}$
for radial scalar functions $\varphi ({\bf r})=\varphi (r)$
coincides with the suggested 
scalar Laplacian $\, ^S\Delta^{D}_r$ for this function,
\be \label{NI-R-2}
^{St}\Delta^{D} \varphi (r) = \, ^S\Delta^{D} \varphi (r) .
\ee
The Stillinger's Laplacian can be applied 
only for scalar fields and it cannot be used 
to describe vector fields ${\bf u}=u_r(r) \, {\bf e}_r$ 
because Stillinger's Laplacian for $D=3$ is not equal to 
the usual vector Laplacian for $\mathbb{R}^3$,
\be
^{St}\Delta^{3} {\bf u}(r) \ne \, \Delta {\bf u}(r) =
\Bigl( 
\frac{\partial^2 u_r}{\partial r^2} + \frac{2}{r} \, 
\frac{\partial u_r}{\partial r}  -  \frac{2}{r^2} \, u_r 
\Bigr) \, {\bf e}_r  .
\ee
For the vector fields ${\bf u}=u_r(r) \, {\bf e}_r$, 
we should use the vector Laplace operators 
(\ref{V-Delta-D}) and (\ref{V-Delta-D}).

Note that the gradient, divergence, curl operator and vector Laplacian are not considered in Stillinger's paper \cite{Stillinger}


\section{Non-integer dimensional space for complex fractal media
with $d \ne D-1$}


In general, the dimension $D$ of the ball region $B_D$
of fractal materials and the dimension $d$ of boundary 
$S_d=\partial B_D$ of this region
are not related by the equation $d = D-1$, i.e.,
\be
\operatorname{dim} (\partial B_D) \ne 
\operatorname{dim} (B_D) -1 , 
\ee
where $\operatorname{dim} (B_D)=D$.
We assume that dimension of the boundary $S_d= \partial B_D $ is
\be
\operatorname{dim} (S_d) = d .
\ee

Considering an infinitely thin hollow ball $B_D$
with $R_1=r$ and $R_2=r+ \Delta r$, we can use
\be
\operatorname{Div}^{D,d}_{r} {\bf u} =
\lim_{ \Delta r \to 0}
\frac{ \Phi_{\bf u} ( S_d (r;r+\Delta r) ) }{
V(B_D(r;r+\Delta r))} 
\ee
in order to derive an expression for the divergence 
for the case $d \ne D-1$.

The flux of the vector field ${\bf u}=u_r(r) \, {\bf e}_r$ 
across a surface $S_d$ is
\[ 
\Phi_{\bf u} ( S_d (r;r+\Delta r) ) =
\frac{2 \, \pi^{(d+1)/2}}{\Gamma((d+1)/2)} 
\, \Bigl( u_r(r+\Delta r) \, (r+\Delta r)^{d}
- u_r(r) \, r^{d} \Bigr) =
\]
\be \label{Flux-1d}
= \frac{2 \, \pi^{(d+1)/2}}{\Gamma((d+1)/2)} 
\, \Bigl( r^{d} \, \frac{\partial u_r(r)}{\partial r} +
d \, r^{d-1} \, u_r (r) +O(\Delta r) \Bigr) \Delta r .
\ee

Using (\ref{Vol-1}) and (\ref{Flux-1d}), we get
\be
\frac{\Phi_{\bf u} (S_d(r;r+\Delta r)) }{ 
V(B_D(r;r+\Delta r))}  =
\frac{\pi^{(d+1-D)/2} \, \Gamma (D/2) }{\Gamma ((d+1)/2)}
\Bigl( \frac{1}{r^{D-1-d}}
\frac{\partial u_r(r)}{\partial r} + \frac{d}{r^{D-d}} \, u_r(r) \Bigr) +O(\Delta r)  .
\ee
Using the limit $\Delta r \to 0$, we get
the divergence operator for fractal media with $d \ne D-1$ 
in the form
\be \label{Div-Dd}
\operatorname{Div}^{D,d}_{r} {\bf u} =\frac{\pi^{(d+1-D)/2} \, \Gamma (D/2) }{\Gamma ((d+1)/2)}
\left( \frac{1}{r^{D-1-d}}
\frac{\partial u_r(r)}{\partial r} + \frac{d}{r^{D-d}} \, u_r(r) 
\right) .
\ee
We can define the parameter
\be \label{ar}
\alpha_r = D-d ,
\ee
that can be interpreted as a dimension of medium 
along the radial direction. Using (\ref{ar}), 
equation (\ref{Div-Dd}) can be rewritten in the form
\be \label{Div-Dd2}
\operatorname{Div}^{D,d}_{r} {\bf u} = 
\pi^{(1-\alpha_r)/2} \, 
\frac{\Gamma ((d+\alpha_r)/2) }{ \Gamma ((d+1)/2)}
\left( \frac{1}{r^{\alpha_r-1}}
\frac{\partial u_r(r)}{\partial r} + 
\frac{d}{r^{\alpha_r}} \, u_r(r) \right) .
\ee
This is divergence operator for non-integer dimensional 
continuum models of fractal materials with $d \ne D-1$.
For $\alpha_r=1$, i.e. $d=D-1$, equations 
(\ref{Div-Dd}) and (\ref{Div-Dd2}) give (\ref{Div-D}).


We can assume that the gradient for the scalar field
$\varphi({\bf r})= \varphi(r)$ 
may depend on the radial dimension $\alpha_r$
in the form
\be \label{Grad-Dd}
\operatorname{Grad}^{D,d}_{r} \varphi = 
\frac{\Gamma (\alpha_r/2)}{ \pi^{\alpha_r/2} \, r^{\alpha_r-1}} \,
\frac{\partial \varphi(r)}{\partial r} \, {\bf e}_r ,
\ee
because expression (\ref{Grad-Dd}) 
can be represented by the equation 
\be \label{Grad-Dd2}
\operatorname{Grad}^{D,d}_{r} \varphi =
\lim_{ \Delta r \to 0}
\frac{ 2( \varphi (r+\Delta r) -\varphi (r))}{
V(B_{\alpha_r}(r;r+\Delta r))} ,
\ee
where 
\be V(B_{\alpha_r}(r;r+\Delta r)) =  
\frac{2 \, \pi^{\alpha_r/2}}{\Gamma(\alpha_r/2)} 
\, \Bigl( r^{\alpha_r-1} \, 
\Delta r +O((\Delta r)^2) \Bigr) . \ee 
The presence of the factor of $2$ in (\ref{Grad-Dd2}) 
is due to the fact that 
for $D=1$, $r$ is integrated from $-R$ to $R$, 
and when the limits are taken as $0$ and $R$, 
one gets a factor of $2$.

For $\alpha_r=1$, equation (\ref{Grad-Dd}) gives (\ref{Grad-D}).

Using the operators (\ref{Grad-Dd}) and (\ref{Div-Dd})
for the fields $\varphi=\varphi (r)$ and 
${\bf u}=u(r) \, {\bf e}_r$,
we can get the scalar and vector Laplace operators 
for the case $d \ne D-1$ by the equation
\be
^S\Delta^{D,d}_r \varphi= 
\operatorname{Div}^{D,d}_r 
\operatorname{Grad}^{D,d}_{r} \varphi , \quad
^V\Delta^{D,d}_r {\bf u} = 
\operatorname{Grad}^{D,d}_r 
\operatorname{Div}^{D,d}_{r} {\bf u} .
\ee
Then the scalar Laplacian for $d \ne D-1$ 
for the field $\varphi=\varphi (r)$ is
\be \label{S-Delta-Dd}
^S\Delta^{D,d}_r \varphi= 
\frac{\Gamma ((d+\alpha_r)/2) \, \Gamma (\alpha_r/2)}{ 
\pi^{\alpha_r-1/2} \, \Gamma ((d+1)/2)}
\Bigl(  \frac{1}{r^{2 \alpha_r-2}} \, 
\frac{\partial^2 \varphi}{\partial r^2} + 
\frac{d+1-\alpha_r}{r^{2\alpha_r-1}} \, 
\frac{\partial \varphi}{\partial r} \Bigr) ,
\ee
and the vector Laplacian for $d \ne D-1$  
for the field ${\bf u}=u(r) \, {\bf e}_r$ is
\be \label{V-Delta-Dd}
^V\Delta^{D,d}_r {\bf u} = 
\frac{\Gamma ((d+\alpha_r)/2) \, \Gamma (\alpha_r/2)}{ 
\pi^{\alpha_r-1/2} \, \Gamma ((d+1)/2)}
\Bigl( \frac{1}{r^{2 \alpha_r-2}} \, 
\frac{\partial^2  u_r }{\partial r^2} 
+ \frac{d+1-\alpha_r}{r^{2\alpha_r-1}} \, 
\frac{\partial  u_r }{\partial r} 
- \frac{d \alpha_r}{r^{2\alpha_r}} \,  u_r 
\Bigr) \, {\bf e}_r .
\ee

The vector differential operators (\ref{Grad-Dd}), 
(\ref{Div-Dd}), (\ref{S-Delta-Dd}) and (\ref{V-Delta-Dd})
allow us to describe complex fractal materials with 
the boundary dimension of the regions $d \ne D-1$.

\section{Applications in mechanics and physics}

In this section, we consider some applications
of vector calculus in non-integer dimensional space
to the elasticity theory, the heat processes, and electrodynamics.

\subsection{Elasticity of fractal material}

For homogenous and isotropic materials, the equation 
of linear elasticity \cite{LL,Hahn}
for the displacement vector fields ${\bf u} = {\bf u}({\bf r},t)$ 
has the form
\be \label{EL-1}
\lambda \, \operatorname{grad} \operatorname{div} {\bf u}
 + 2 \mu \, \Delta \, {\bf u} + {\bf f} = \rho \, D^2_t {\bf u} ,
\ee
where $\lambda$ and $\mu$ are the Lame coefficients,
and ${\bf f}$ is the external force density vector field.

If the deformation in the material is described by 
${\bf u}({\bf r},t) = u(r,t) \, {\bf e}_r$, then
equation (\ref{EL-1}) has the form
\be \label{EL-2}
(\lambda  + 2 \mu ) \, \Delta \, {\bf u}(r,t) + {\bf f}(r,t) = 
\rho \, D^2_t {\bf u} (r,t) .
\ee

A generalization of equations (\ref{EL-2}) 
for fractal material in the framework of
non-integer dimensional models  has the form
\be \label{F-1}
(\lambda+2 \mu ) \, ^V\Delta^{D}_r \,{\bf u}(r,t) +{\bf f}(r,t) = 
\rho \, D^2_t {\bf u} (r,t) ,
\ee
where $^V\Delta^{D}_r$ is defined by (\ref{V-Delta-D}).
Equation (\ref{F-1}) describes dynamics of
displacement vector for fractal materials. 
For static case, equation (\ref{F-1}) has the form
\be \label{F-2}
^V\Delta^{D}_r \, {\bf u}(r) + 
(\lambda+2 \mu )^{-1} \,{\bf f}(r) = 0 ,
\ee
where ${\bf u}= u_r \, {\bf e}_r$ and 
${\bf f}= f (r)\, {\bf e}_r$. 

Let us consider some two problems for elasticity
of fractal materials.


\subsubsection*{Elasticity of fractal hollow ball 
with pressure inside and outside}

Let us determine the deformation of a hollow fractal ball 
with internal radius $R_1$ and external radius $R_2$,
with the pressure $p_1$ inside and the pressure $p_2$ outside.

We can use the spherical polar coordinates
with the origin at the center of the ball.
The displacement vector ${\bf u}$ is everywhere radial,
and it is a function of $r=|{\bf r}|$ alone.
Then the equilibrium equation for fractal ball is
\be \label{F-2f0}
(\lambda+2 \mu ) \ ^V\Delta^{D}_r \, {\bf u}(r) = 0 ,
\ee
where ${\bf u}= u_r \, {\bf e}_r$. 
Using (\ref{V-Delta-D}), we represent equation (\ref{F-2f0}) 
in the form
\be \label{F-3f0}
\frac{\partial^2 u_r(r)}{\partial r^2} +
\frac{D-1}{r} \, \frac{\partial u_r(r)}{\partial r} -
\frac{D-1}{r^2} \, u_r(r) = 0.
\ee
The solution of (\ref{F-3f0}) is
\be
u(r) = C_1 \, r + C_2 \, r^{1-D}  .
\ee
The constants $C_1$ and $C_2$ are determined from  
the boundary conditions for radial stress
\be
\sigma_{rr} (R_1) = - p_1 , \quad \sigma_{rr} (R_2) = - p_2 .
\ee
Using that the radial components of the stress is
\be
\sigma_{rr}(r) = 
(2\, \mu + \lambda) \, \frac{\partial u_r}{\partial r} + 
\lambda \, \frac{D-1}{r} \, u_r  ,
\ee
we get
\be
C_1 =
 \frac{-(p_2 \,  R^{D}_2 - p_1 \,  R^{D}_1) }{(2\, \mu + D\, \lambda) \, (R^{D}_2 -R^{D}_1)} ,
\ee
\be
C_2 = \frac{p_2 - p_1}{ 2 \, (1-D) \, \mu \, 
(R^{D}_2 -R^{D}_1)} .
\ee
Then the radial components of the stress is
\be \label{sigma-1}
\sigma_{rr}(r) = 
\frac{-(p_2 \,  R^{D}_2 - p_1 \,  R^{D}_1) }{R^{D}_2 -R^{D}_1} 
+ \frac{(p_2 - p_1) \, (R_1 \, R_2)^D }{R^{D}_2 -R^{D}_1} \, r^{-D} .
\ee
The stress distribution in a ball with pressure $p_1=p$ inside
and $p_2=0$ outside is gives by
\be
\sigma_{rr}(r) =  \frac{ p \,  R^{D}_1 }{R^{D}_2 -R^{D}_1}
\left( 1 - \left(\frac{R_2}{r}\right)^D \right) .
\ee

The stress distribution in an infinite elastic medium with spherical cavity with radius $R$ subjected to hydrostatic
compression is
\be
\sigma_{rr}(r) = -p \, \left( 1 - \left(\frac{R}{r}\right)^D \right) 
\ee
that can be obtained by putting $R_1=R$, $R_2 \to \infty$,
$p_1=0$ and $p_2=p$ in equation (\ref{sigma-1}).

\subsubsection*{Elasticity of cylindrical fractal solid pipe 
with pressure inside and outside}

Let us consider the deformation of a fractal solid 
cylindrical pipe with internal radius $R_1$ 
and external radius $R_2$ 
with pressure $p_1$ a inside and pressure $p_2$ outside.
We use the cylindrical coordinates with the $z$-axis
along the axis of the pipe. 
When the pressure is uniform along the pipe,
the deformation is a purely radial displacement 
${\bf u}=u_r(r) \, {\bf e}_r$, where ${\bf e}_r={\bf r}/r$.
The equation for the displacement $u_r(r)$ 
in fractal pine is
\be \label{Cyl-1.1}
\frac{\partial^2 u_r(r)}{\partial r^2} + \frac{D-2}{r} \, 
\frac{\partial u_r}{\partial r} - \frac{D-2}{r^2} \, u_r = 0 ,
\ee
where $0<D \le 3$. 
If $D=3$, we get the usual (non-fractal) case.

The general solution of equation (\ref{Cyl-1.1}),  
where $D \ne 1$, $D \ne 2$, has the form
\be \label{Sol-P-1}
u_r(r) = C_1\, r + C_2\, r^{2-D} .
\ee
Equations (\ref{Cyl-1.1}) with $D=1$ has the general solution
\be
u_r(r) = C_1\, r + C_2\, r \, \ln (r) .
\ee
For $D=2$, equations (\ref{Cyl-1.1}) has the solution
\be
u_r(r) = C_1\, + C_2\, r .
\ee
Note that dimensions $D=1$ or $D=2$ of the fractal pipe material 
do not correspond 
to the distribution of matter along the line and surface.
These dimensions describe a distribution of matter
in 3-dimensional space (in the volume of pipe) 
such that the mass dimensions are equal to $D$. 

The constants $C_1$ and $C_2$ are determined by
boundary conditions.
Using that pressure $p_1$ a inside and pressure $p_2$ outside, 
we get the boundary condition in the form
\be \label{BCond-1}
\sigma_{rr} (R_1) = - \, p_1 , \quad
\sigma_{rr} (R_2) = - \, p_2 .
\ee
Using (\ref{Sol-P-1}) and
\be
\sigma_{rr} =(2\mu+\lambda) \frac{\partial u_r}{\partial r} +
\lambda \frac{D-2}{r} u_r = (2 \mu+ \lambda\, (D-1)) \, C_1 
- 2 \, \mu \, (D-2)\, C_2\, r^{1-D} ,
\ee
the boundary condition (\ref{BCond-1}) gives
\be
C_1 = - \frac{ p_1 \, R^{1-D}_2 - p_2 \, R^{1-D}_1 }{ 
(2 \mu+ \lambda\, (D-1)) \, ( R^{1-D}_2 - R^{1-D}_1) } ,
\ee
\be
C_2 = \frac{ p_2  - p_1 }{2 \, \mu \, (D-2) 
\, (R^{1-D}_2 - R^{1-D}_1) } .
\ee
The stress for $2<D<3$ or $1<D<2$ can represented in the form
\be \label{stress-rr-2}
\sigma_{rr} = \frac{ p_1 \, R^{D-1}_1 - p_2 \, R^{D-1}_2 }{ 
( R^{D-1}_2 - R^{D-1}_1) }  
- \frac{ p_2  - p_1 }{ (R^{D-1}_2 - R^{D-1}_1) } \, 
\left( \frac{R_1\, R_2}{r} \right)^{D-1} .
\ee
For the boundary conditions 
$\sigma_{rr}(R_2)=0$ and $\sigma_{rr}(R_1)=-p$,
i.e. $p_2=0$ and $p_1=p$ for (\ref{stress-rr-2}),
we have the solution 
\be \label{stress-rr-3}
\sigma_{rr} = \frac{ p \, R^{D-1}_1}{ 
( R^{D-1}_2 - R^{D-1}_1) }  \left( 1 -
\left( \frac{R_2}{r} \right)^{D-1} \right) .
\ee
This is the deformation of cylindrical pipe
with a pressure $p$ inside and no pressure outside.
For $D=3$, equation (\ref{stress-rr-3}) 
has the well-known form
\be \label{stress-rr-4}
\sigma_{rr} = \frac{ p \, R^{2}_1}{ 
( R^{2}_2 - R^{2}_1) }  \left( 1 -
\left( \frac{R_2}{r} \right)^{2} \right) 
\ee
that describes the stress of non-fractal material of pipe.

\subsection{Heat equation for fractal materials}

The heat equation is
\be \label{Heat-3}
\frac{\partial \varphi ({\bf r},t)}{\partial t} - 
a \, \Delta \varphi ({\bf r},t) = 
\frac{1}{c_p \, \rho} \, q({\bf r},t) ,
\ee
where $\varphi ({\bf r},t)$ is the heat density of a medium,
$q({\bf r},t)$ is the heat source density,
and $a$ is the thermal diffusivity 
$a= {k}/{c_p \, \rho}$, where $k$ is thermal conductivity, 
$\rho$ is density, $c_p$ is specific heat capacity.

A generalization of equation (\ref{Heat-3}) 
for fractal material, has the form of the heat equation 
in the non-integer dimensional space
\be
\frac{\partial \varphi (r,t)}{\partial t} - 
a \, ^S\Delta^{D}_r \varphi (r,t) = \frac{1}{c_p \, \rho} \, q(r,t)  ,
\ee
where we assume that the fields $\varphi (r,t)$ 
and $q (r,t)$ are not depend on the angles.

Using (\ref{S-Delta-D}), we get the following equations 
for the ball
\be \label{FHE-1a}
\frac{\partial^2 \varphi (r,t)}{\partial r^2} + 
\frac{D-1}{r} \, \frac{\partial \varphi (r,t)}{\partial r} + 
\frac{1}{c_p \, \rho} \, q(r,t) =
\frac{1}{a} \frac{\partial \varphi (r,t)}{\partial t}  .
\ee
For the pipe and cylinder, we get 
\be \label{FHE-1b}
\frac{\partial^2 \varphi (r,t)}{\partial r^2} + 
\frac{D-2}{r} \, \frac{\partial \varphi (r,t)}{\partial r} + 
\frac{1}{c_p \, \rho} \, q(r,t) =
\frac{1}{a} \frac{\partial \varphi (r,t)}{\partial t}  .
\ee
Steady states is described by the equation
\be \label{FHE-2}
\frac{\partial^2 \varphi (r)}{\partial r^2} + 
\frac{D-1}{r} \, \frac{\partial \varphi (r)}{\partial r} + 
\frac{1}{c_p \, \rho} \, q(r) = 0 .
\ee
The general solution of equation (\ref{FHE-2}) is
\be \label{gsol1}
\varphi (r) = C_1 + C_2 \, r^{2-D} +
\frac{1}{c_p \, \rho \, (D-2)}
\left( r^{2-D} \, \int q(r) \, r^{D-1} \, dr - 
\int q(r) \, r \, dr \right) ,
\ee
where the constants $C_1$ and $C_2$ are determined by the boundary condition.
For the constant heat source density 
$q(r)=q_0=\operatorname{const}$, equation (\ref{gsol1})
has the form
\be \label{gsol2}
\varphi (r) = C_1 + C_2 \, r^{2-D} -
\frac{q_0}{2 \, D\, c_p \, \rho} r^2 .
\ee
For $D=3$, we get the well-known equation for non-fractal material.


\subsection{Electric field of fractal charged infinite cylinder}

Let us consider a uniformly fractal charged 
infinite circular cylinder
of radius $R$ with a volume charge density 
$\rho = \operatorname{const}$
and non-integer dimension $2 < D \le 3$.
Using the Poisson equation for scalar potential 
created by an infinite circular cylinder. 
We assume that the $Z$-axis is directed along 
the axis of the cylinder. 
Due to the axial symmetry of the charge distribution 
the potential is also axially symmetric.
Therefore it is convenient to use 
a cylindrical coordinate system.
The Poisson equation for scalar field $\varphi(r)$
in non-integer dimensional space has the form
\be \label{Elec-1}
\frac{\partial^2 \varphi}{\partial r^2} + \frac{D-2}{r} \, 
\frac{\partial \varphi}{\partial r} = 
- \frac{\rho}{\varepsilon_0}
\quad (0<r<R) ,
\ee 
\be \label{Elec-2}
\frac{\partial^2 \varphi}{\partial r^2} + \frac{D-2}{r} \, 
\frac{\partial \varphi}{\partial r} = 0
\quad (r>R) .
\ee
The general solution of equations 
(\ref{Elec-1}) and (\ref{Elec-2}) are
\be \label{Elec-3}
\varphi(r) = C_1 + C_2 \, r^{3-D} - 
\frac{\rho \, r^2}{2 \, \varepsilon_0 \, (D-1)}
\quad (0<r<R) ,
\ee 
\be \label{Elec-4}
\varphi(r) = C_3 + C_4 \, r^{3-D} 
\quad (r>R) ,
\ee
where $C_1$, $C_2$, $C_3$, $C_4$ are the integration constants,
and $2<D \le 3$.
For the case $D=3$, the general solution of equations 
(\ref{Elec-1}) and (\ref{Elec-2}) has the well-known form
\be \label{Elec-3D3}
\varphi(r) =  C_1+ C_2 \, \ln (r)
- \frac{ \rho \, r^2}{4 \, \varepsilon_0}
\quad (0<r<R) ,
\ee 
\be \label{Elec-4D3}
\varphi(r) =  C_3+ C_4 \, \ln (r) \quad (r>R) .
\ee
The electric fields 
\be \label{Elec-5}
{\bf E}(r) = - \operatorname{Grad}^{D}_r \, \varphi = 
- \frac{\partial \varphi(r)}{\partial r} \, {\bf e}_r 
\ee
for potentials  (\ref{Elec-3}) and (\ref{Elec-4}) are
\be \label{Elec-3E}
{\bf E}(r) = \Bigl( (D-3) \, C_2 \, r^{2-D} +
\frac{\rho \, r}{\varepsilon_0 \, (D-1)} \Bigr) \, {\bf e}_r 
\quad (0<r<R) ,
\ee 
\be \label{Elec-4E}
{\bf E}(r)  = (D-3) \, C_4 \, r^{2-D} \, {\bf e}_r 
\quad (r>R) .
\ee
Because the electric field (\ref{Elec-5})
must be finite at all points, and 
$r^{2-D} \to \infty $ for $r \to 0$ and $2<D \le 3$, 
it is necessary put $C_2 = 0$.
Conveniently potential normalized by the condition 
$\varphi(0)=0$, then we get $C_1 =0$.
Because there are no surface charges, 
then the electric field (\ref{Elec-5})
at the surface of the cylinder $r=R$ is continuous, i.e. 
the derivative of the potential should be continuous. 
The conditions of continuity of the potential and its derivative at $r=R$ 
give two algebraic equations that allow us to determine 
the remaining two constants $C_3$ and $C_4$ by the equations
\be \label{Elec-6}
- \frac{\rho \, R^2}{2 \, \varepsilon_0 \, (D-1)}
= C_3 + C_4 \, R^{3-D} ,
\ee
\be \label{Elec-7}
\frac{\rho \, R}{\varepsilon_0 \, (D-1)} = 
(D-3) \, C_4 \, R^{2-D} .
\ee
Then we have
\be \label{Elec-8}
C_3 = - \frac{\rho \, R^2}{2 \, \varepsilon_0 \, (D-3)}
, \quad
C_4 = \frac{\rho \, R^{D-1}}{\varepsilon_0 \, (D-1) \, (D-3)} .
\ee
As a result, the potential is
\be \label{Elec-9}
\varphi(r) = - \frac{\rho \, r^2}{2 \, \varepsilon_0 \, (D-1)}
\quad (0<r\le R) ,
\ee 
\be \label{Elec-10}
\varphi(r) = - \frac{\rho \, R^2}{2 \, \varepsilon_0 \, (D-3)}
+ \frac{\rho \, R^{D-1}}{\varepsilon_0 \, (D-1) \, (D-3)} 
\, r^{3-D} \quad (r \ge R) .
\ee
Using (\ref{Elec-5}) and (\ref{Elec-9} - \ref{Elec-10}), 
the electric field has the form
\be \label{Elec-11}
{\bf E}(r) = \frac{\rho \, r}{\varepsilon_0 \, (D-1)} 
\, {\bf e}_r \quad (0 < r \le R) ,
\ee 
\be \label{Elec-12}
{\bf E}(r) = \frac{\rho \, R^{D-1} \, r^{2-D}}{ \varepsilon_0 \, (D-1)} 
 \, {\bf e}_r 
\quad (r \ge R) .
\ee
For $D=3$, we get the well-known results of non-fractal case.

Equation (\ref{Elec-11}) 
can be represented in the form
\be \label{Elec-11e}
{\bf E}(r) = 
\frac{\rho \, r}{2 \varepsilon_0 \varepsilon_{eff, in} } 
\, {\bf e}_r \quad (0 < r \le R) ,
\ee 
where $\varepsilon_{eff, in} =(D-1)/2$ is an effective permittivity
of fractal materials.
Consider the charge per unit length
\be \label{109}
\tau_D = \rho \, V_{D-1} =
\rho \, \frac{\pi^{(D-1)/2} R^{D-1}}{ \Gamma((D+1)/2)} .
\ee
For $D=3$, equation (\ref{109})
gives the value $ \tau_3 = \rho \, \pi \, R^2$
for non-fractal charge cylinder.
Using (\ref{109}) equation (\ref{Elec-12})
can be represented in the form
\be \label{Elec-12e}
{\bf E}(r) = 
\frac{1}{2 \, \pi \, \varepsilon_0 \, \varepsilon_{eff, out} } 
\, \frac{\tau_{D}}{r^{D-2}} \, {\bf e}_r 
\quad (r \ge R) ,
\ee
where the effective permittivity
\be
\varepsilon_{eff, out} = \frac{(D-1)}{ 
2 \, \, \pi^{(3-D)/2} \, \Gamma( (D-1)/2 )} .
\ee

The electric field in the fractal homogeneous charged cylinder 
is analogous to the non-fractal case up to the factor $\varepsilon_{eff, in}$. We have
a linear dependence on the distance from the cylinder axis
for $0 < r \le R$.
Electric field outside the fractal charged cylinder 
differs from non-fractal case.
For $r \ge R$, we have
power-law dependence on the distance from the cylinder axis.
In addition
the electric field outside the cylinder is reduced by  
the effective permittivity $\varepsilon_{eff, out}$.

\section{Conclusion}

In this paper, differential operators of vector calculus
for non-integer dimensional space is suggested
to describe fractal media and materials
in the framework of continuum models.
The first and second order differential operators 
for non-integer dimensional space
are proposed for rotationally covariant scalar 
and vector functions.
We consider some applications for the case 
of spherical and axial symmetries of 
the fractal material.
Elasticity of fractal hollow ball and 
fractal cylindrical pipe,
heat distribution in fractal media,
and electrostatic field of fractal charged cylinder
are described to illustrate the suggested approach.
In general, we can consider not only 
first and second order differential operators 
for non-integer dimensional space.
The differential and integral operators of fractional orders 
can also be considered for non-integer dimensional spaces
to take into account non-locality of materials.
We can note that a dimensional continuation of
the Riesz fractional integrals and derivatives \cite{KST,FC2}
to generalize differential and integrals of fractional 
orders for non-integer dimensional space 
has been considered in \cite{MA2010}.



\end{document}